\input amssym.def
\input epsf
\epsfclipon


\magnification=\magstephalf
\hsize=14.0 true cm
\vsize=19 true cm
\hoffset=1.0 true cm
\voffset=2.0 true cm

\abovedisplayskip=12pt plus 3pt minus 3pt
\belowdisplayskip=12pt plus 3pt minus 3pt
\parindent=1.0em


\font\sixrm=cmr6
\font\eightrm=cmr8
\font\ninerm=cmr9

\font\sixi=cmmi6
\font\eighti=cmmi8
\font\ninei=cmmi9

\font\sixsy=cmsy6
\font\eightsy=cmsy8
\font\ninesy=cmsy9

\font\sixbf=cmbx6
\font\eightbf=cmbx8
\font\ninebf=cmbx9

\font\eightit=cmti8
\font\nineit=cmti9

\font\eightsl=cmsl8
\font\ninesl=cmsl9

\font\sixss=cmss8 at 8 true pt
\font\sevenss=cmss9 at 9 true pt
\font\eightss=cmss8
\font\niness=cmss9
\font\tenss=cmss10

 at 12 true pt
\font\bigrm=cmr10 at 12 true pt
 at 12 true pt

 at 14 true pt
\font\Bigrm=cmr12 at 16 true pt
 at 14 true pt

\catcode`@=11
\newfam\ssfam

\def\tenpoint{\def\rm{\fam0\tenrm}%
    \textfont0=\tenrm \scriptfont0=\sevenrm \scriptscriptfont0=\fiverm
    \textfont1=\teni  \scriptfont1=\seveni  \scriptscriptfont1=\fivei
    \textfont2=\tensy \scriptfont2=\sevensy \scriptscriptfont2=\fivesy
    \textfont3=\tenex \scriptfont3=\tenex   \scriptscriptfont3=\tenex
    \textfont\itfam=\tenit                  \def\it{\fam\itfam\tenit}%
    \textfont\slfam=\tensl                  \def\sl{\fam\slfam\tensl}%
    \textfont\bffam=\tenbf \scriptfont\bffam=\sevenbf
    \scriptscriptfont\bffam=\fivebf
                                            \def\bf{\fam\bffam\tenbf}%
    \textfont\ssfam=\tenss \scriptfont\ssfam=\sevenss
    \scriptscriptfont\ssfam=\sevenss
                                            \def\ss{\fam\ssfam\tenss}%
    \normalbaselineskip=13pt
    \setbox\strutbox=\hbox{\vrule height8.5pt depth3.5pt width0pt}%
    \let\big=\tenbig
    \normalbaselines\rm}

\def\ninepoint{\def\rm{\fam0\ninerm}%
    \textfont0=\ninerm      \scriptfont0=\sixrm
                            \scriptscriptfont0=\fiverm
    \textfont1=\ninei       \scriptfont1=\sixi
                            \scriptscriptfont1=\fivei
    \textfont2=\ninesy      \scriptfont2=\sixsy
                            \scriptscriptfont2=\fivesy
    \textfont3=\tenex       \scriptfont3=\tenex
                            \scriptscriptfont3=\tenex
    \textfont\itfam=\nineit \def\it{\fam\itfam\nineit}%
    \textfont\slfam=\ninesl \def\sl{\fam\slfam\ninesl}%
    \textfont\bffam=\ninebf \scriptfont\bffam=\sixbf
                            \scriptscriptfont\bffam=\fivebf
                            \def\bf{\fam\bffam\ninebf}%
    \textfont\ssfam=\niness \scriptfont\ssfam=\sixss
                            \scriptscriptfont\ssfam=\sixss
                            \def\ss{\fam\ssfam\niness}%
    \normalbaselineskip=12pt
    \setbox\strutbox=\hbox{\vrule height8.0pt depth3.0pt width0pt}%
    \let\big=\ninebig
    \normalbaselines\rm}

\def\eightpoint{\def\rm{\fam0\eightrm}%
    \textfont0=\eightrm      \scriptfont0=\sixrm
                             \scriptscriptfont0=\fiverm
    \textfont1=\eighti       \scriptfont1=\sixi
                             \scriptscriptfont1=\fivei
    \textfont2=\eightsy      \scriptfont2=\sixsy
                             \scriptscriptfont2=\fivesy
    \textfont3=\tenex        \scriptfont3=\tenex
                             \scriptscriptfont3=\tenex
    \textfont\itfam=\eightit \def\it{\fam\itfam\eightit}%
    \textfont\slfam=\eightsl \def\sl{\fam\slfam\eightsl}%
    \textfont\bffam=\eightbf \scriptfont\bffam=\sixbf
                             \scriptscriptfont\bffam=\fivebf
                             \def\bf{\fam\bffam\eightbf}%
    \textfont\ssfam=\eightss \scriptfont\ssfam=\sixss
                             \scriptscriptfont\ssfam=\sixss
                             \def\ss{\fam\ssfam\eightss}%
    \normalbaselineskip=10pt
    \setbox\strutbox=\hbox{\vrule height7.0pt depth2.0pt width0pt}%
    \let\big=\eightbig
    \normalbaselines\rm}

\def\tenbig#1{{\hbox{$\left#1\vbox to8.5pt{}\right.\n@space$}}}
\def\ninebig#1{{\hbox{$\textfont0=\tenrm\textfont2=\tensy
                       \left#1\vbox to7.25pt{}\right.\n@space$}}}
\def\eightbig#1{{\hbox{$\textfont0=\ninerm\textfont2=\ninesy
                       \left#1\vbox to6.5pt{}\right.\n@space$}}}

\font\sectionfont=cmbx10
\font\subsectionfont=cmti10

\def\figurecaptionfont{\ninepoint}
\def\tablecaptionfont{\ninepoint}
\def\footnotefont{\eightpoint}


\newcount\equationno
\newcount\bibitemno
\newcount\figureno
\newcount\tableno

\equationno=0
\bibitemno=0
\figureno=0
\tableno=0


\footline={\ifnum\pageno=0{\hfil}\else
{\hss\rm\the\pageno\hss}\fi}


\def\section #1. #2 \par
{\vskip0pt plus .10\vsize\penalty-100 \vskip0pt plus-.10\vsize
\vskip 1.6 true cm plus 0.2 true cm minus 0.2 true cm
\global\def\equationlabel{#1}
\global\equationno=0
\leftline{\sectionfont #1. #2}\par
\immediate\write\terminal{Section #1. #2}
\vskip 0.7 true cm plus 0.1 true cm minus 0.1 true cm
\noindent}


\def\subsection #1 \par
{\vskip0pt plus 0.8 true cm\penalty-50 \vskip0pt plus-0.8 true cm
\vskip2.5ex plus 0.1ex minus 0.1ex
\leftline{\subsectionfont #1}\par
\immediate\write\terminal{Subsection #1}
\vskip1.0ex plus 0.1ex minus 0.1ex
\noindent}


\def\appendix #1. #2 \par
{\vskip0pt plus .20\vsize\penalty-100 \vskip0pt plus-.20\vsize
\vskip 1.6 true cm plus 0.2 true cm minus 0.2 true cm
\global\def\equationlabel{\hbox{\rm#1}}
\global\equationno=0
\leftline{\sectionfont Appendix #1. #2}\par
\immediate\write\terminal{Appendix #1. #2}
\vskip 0.7 true cm plus 0.1 true cm minus 0.1 true cm
\noindent}



\def\equation#1{$$\displaylines{\qquad #1}$$}
\def\enum{\global\advance\equationno by 1
\hfill\llap{(\equationlabel.\the\equationno)}}
\def\noenum{\hfill}
\def\next#1{\cr\noalign{\vskip#1}\qquad}


\def\ifundefined#1{\expandafter\ifx\csname#1\endcsname\relax}

\def\ref#1{\ifundefined{#1}?\immediate\write\terminal{unknown reference
on page \the\pageno}\else\csname#1\endcsname\fi}

\newwrite\terminal
\newwrite\bibitemlist

\def\bibitem#1#2\par{\global\advance\bibitemno by 1
\immediate\write\bibitemlist{\string\def
\expandafter\string\csname#1\endcsname
{\the\bibitemno}}
\item{[\the\bibitemno]}#2\par}

\def\beginbibliography{
\vskip0pt plus .15\vsize\penalty-100 \vskip0pt plus-.15\vsize
\vskip 1.2 true cm plus 0.2 true cm minus 0.2 true cm
\leftline{\sectionfont References}\par
\immediate\write\terminal{References}
\immediate\openout\bibitemlist=biblist
\frenchspacing\parindent=1.8em
\vskip 0.5 true cm plus 0.1 true cm minus 0.1 true cm}

\def\endbibliography{
\immediate\closeout\bibitemlist
\nonfrenchspacing\parindent=1.0em}


\def\figurecaption#1{\global\advance\figureno by 1
\narrower\figurecaptionfont
Fig.~\the\figureno. #1}

\def\tablecaption#1{\global\advance\tableno by 1
\vbox to 0.5 true cm { }
\centerline{\tablecaptionfont%
Table~\the\tableno. #1}
\vskip-0.4 true cm}

\def\thintablerule{\hrule height0.4pt}

\tenpoint

\immediate\openin\bibitemlist=biblist
\ifeof\bibitemlist\immediate\closein\bibitemlist
\else\immediate\closein\bibitemlist
\input biblist \fi


\def\thismonth{\ifcase\month\or
January\or February\or March\or April\or May\or June\or
July\or August\or September\or October\or November\or December\fi}



\def\rme{{\rm e}}
\def\rmO{{\rm O}}



\def\proof{\noindent{\sl Proof:}\kern0.6em}

\def\frac#1#2{\hbox{$#1\over#2$}}
\def\dual{\mathstrut^*\kern-0.1em}

\def\lvec#1{\setbox0=\hbox{$#1$}
    \setbox1=\hbox{$\scriptstyle\leftarrow$}
    #1\kern-\wd0\smash{
    \raise\ht0\hbox{$\raise1pt\hbox{$\scriptstyle\leftarrow$}$}}
    \kern-\wd1\kern\wd0}
\def\rvec#1{\setbox0=\hbox{$#1$}
    \setbox1=\hbox{$\scriptstyle\rightarrow$}
    #1\kern-\wd0\smash{
    \raise\ht0\hbox{$\raise1pt\hbox{$\scriptstyle\rightarrow$}$}}
    \kern-\wd1\kern\wd0}


\def\nabstar#1{{\nabla\kern0.5pt\smash{\raise 4.5pt\hbox{$\ast$}}
               \kern-5.5pt_{#1}}}

\def\drvstar#1{{\partial\kern0.5pt\smash{\raise 4.5pt\hbox{$\ast$}}
               \kern-6.0pt_{#1}}}

\def\ldrvstar#1{{\lvec{\,\partial}\kern-0.5pt\smash{\raise 4.5pt\hbox{$\ast$}}
               \kern-5.0pt_{#1}}}


\def\MeV{{\rm MeV}}
\def\GeV{{\rm GeV}}

\def\fm{{\rm fm}}




\def\dirac#1{\gamma_{#1}}
\def\diracstar#1#2{
    \setbox0=\hbox{$\gamma$}\setbox1=\hbox{$\gamma_{#1}$}
    \gamma_{#1}\kern-\wd1\kern\wd0
    \smash{\raise4.5pt\hbox{$\scriptstyle#2$}}}


\def\Ad{{\rm Ad}\kern0.1em}


\def\gbar{\bar{g}}
\def\mpi{m_{\pi}}
\def\fpi{f_{\pi}}
\rightline{CERN-TH/2002-321}

\vskip 1.0 true cm 
\centerline{\Bigrm Lattice QCD --- from quark confinement to}
\vskip 1.6ex
\centerline{\Bigrm asymptotic freedom\hskip2pt%
${\vrule height1.5ex depth0.0ex width0.0ex}^{\ast}$
\footnote{}
{\footnotefont%
\hskip-2.5ex$^{\ast}$ Plenary talk, 
International Conference on Theoretical Physics, 
Paris, UNESCO, 22-27 July 2002}
}
\vskip 0.5 true cm
\centerline{\bigrm Martin L\"uscher}
\vskip1ex
\centerline{\it CERN, Theory Division} 
\centerline{\it CH-1211 Geneva 23, Switzerland}

\vskip 0.6 true cm
\thintablerule
\vskip 2.0ex
\ninepoint
\leftline{\bf Abstract}
\vskip 1.0ex\noindent
According to the present understanding, the observed diversity
of the strong interaction phenomena is described
by Quantum Chromodynamics, a gauge field theory with only very 
few parameters.
One of the fundamental questions in this context is 
how precisely the world of mesons and nucleons is related to 
the properties of the theory at high energies,
where quarks and gluons are the important degrees of freedom.
The lattice formulation of QCD combined with
numerical simulations and standard perturbation theory
are the tools that allow one to address this issue
at a quantitative level.
\vskip 2.0ex
\thintablerule

\tenpoint


\section 1. Introduction

Quantum chromodynamics (QCD) is a gauge field theory 
that looks 
rather similar to quantum
electrodynamics. Apart from the fact that
the gauge group is SU(3) instead of U(1),
the lagrange density
\equation{
  {\cal L}_{\rm QCD}=\frac{1}{4}F^a_{\mu\nu}F^a_{\mu\nu}+
  \sum_{q=u,d,s,\ldots}\bar{q}
  \left\{\dirac{\mu}\left(\partial_{\mu}+gA^a_{\mu}T^a\right)+m_q\right\}q
  \enum
}
has the same general form, the first term being
the square of the gauge field tensor and the second a sum over
the contributions of the up, down, strange and the heavy quarks.
No attempt will here be made to explain the detailed structure
of eq.~(1.1), but an important point to note is that
there are no free parameters other than
the gauge coupling $g$ and the
masses $m_u,m_d,\ldots$ of the quarks.

QCD is thus an extremely predictive theory.
It is also difficult to work out
so that only too often ad hoc
approximations need to be made before theory and experiment
can be compared.
Precision tests of QCD are therefore still rare, and 
complex strong interaction phenomena (such as those observed in
heavy ion collisions) will probably never be explained
from first principles.

\subsection 1.1 Quark confinement \& asymptotic freedom

The Feynman rules derived from the QCD lagrangian 
suggest that the quarks are weakly interacting with each
other by exchanging massless vector bosons
(see fig.~1). Similar to an exchange of photons,
the forces that result from this type interaction 
fall off like $1/r^2$ at large distances $r$, and
the energy required to
break up a quark-antiquark bound state is hence finite.
Quarks have, however, never been observed in isolation nor
is there any experimental evidence for massless vector
bosons other than the photon. 

\topinsert
\vbox{
\vskip0.0cm
\epsfxsize=3.3cm\hskip4.65cm\epsfbox{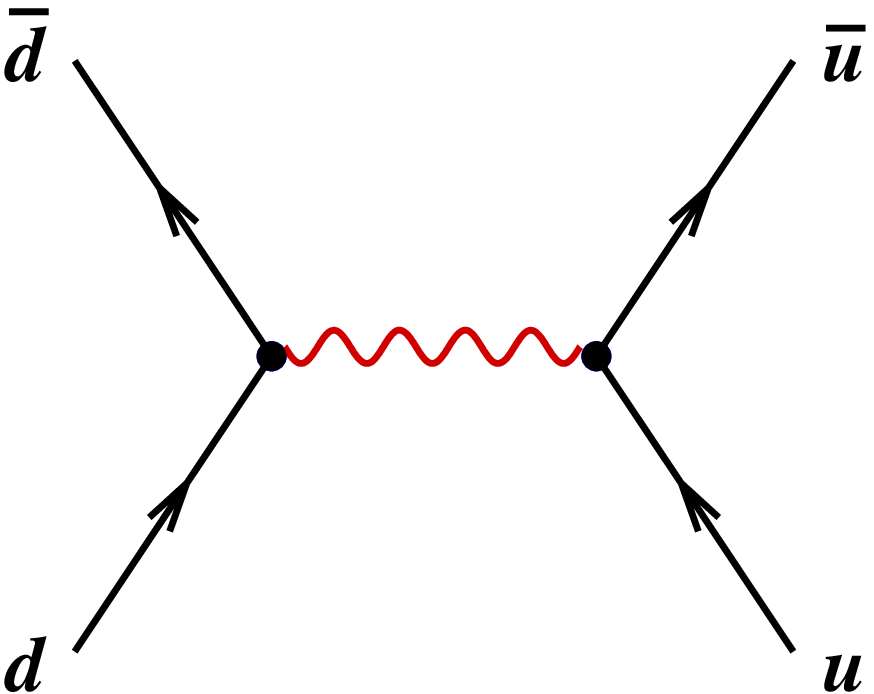}
\vskip0.4cm
\figurecaption{%
In QCD the quarks interact by exchanging 
massless vector bosons (wiggly line) that 
are referred to as gluons. 
They carry SU(3) quantum numbers and couple to all flavours
of quarks 
with equal strength proportional to the gauge coupling $g$.
}
\vskip0.0cm
}
\endinsert

An important hint to the solution of the puzzle
comes from perturbation theory itself.
Once higher-order corrections are included, it turns out that
the strength of the interactions mediated by the vector bosons
depends on the 
magnitude $q$ of the energy-momentum that is transferred between
the quarks. Explicitly, if we introduce 
an effective coupling 
\equation{
  \noenum
  \next{-0.7cm}
  {\kern1.2cm\epsfxsize=1.6cm\epsfbox{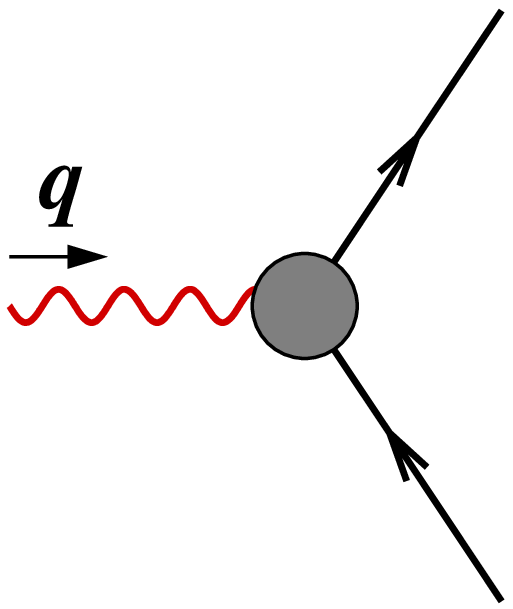}}
  \noenum
  \next{-1.45cm}
  \gbar(q)=
  \enum
  \next{0.2cm}
}
where the shaded circle stands for the sum of all vertex diagrams 
with all possible exchanges of virtual quarks and vector bosons,
it can be shown that
[\ref{GrossWilczek},\ref{Politzer}]
\equation{
   \quad\alpha_{\rm s}(q)\equiv{\gbar(q)^2\over4\pi}
  ={c\over\ln(q/\Lambda)}+\ldots
  \enum
}
for large momenta $q$ and some calculable constant $c$.
The logarithmic decay of the coupling 
(which is referred to as {\it asymptotic freedom}\/) 
is actually observed
in high-energy scattering experiments
(see fig.~2), and from such measurements
the value of the mass scale $\Lambda$ in eq.~(1.3) was
determined to be $213{\hbox{$+38\atop-35$}}\;\MeV$
[\ref{Bethke}]
\footnote{$\dag$}{\footnotefont%
The quoted value of $\Lambda$ 
refers to a particular definition of the effective 
coupling that is known as the ``$\overline{\rm MS}$ scheme
of dimensional regularization".
}.

\topinsert
\vbox{
\vskip0.0cm
\epsfxsize=6.6cm\hskip2.8cm\epsfbox{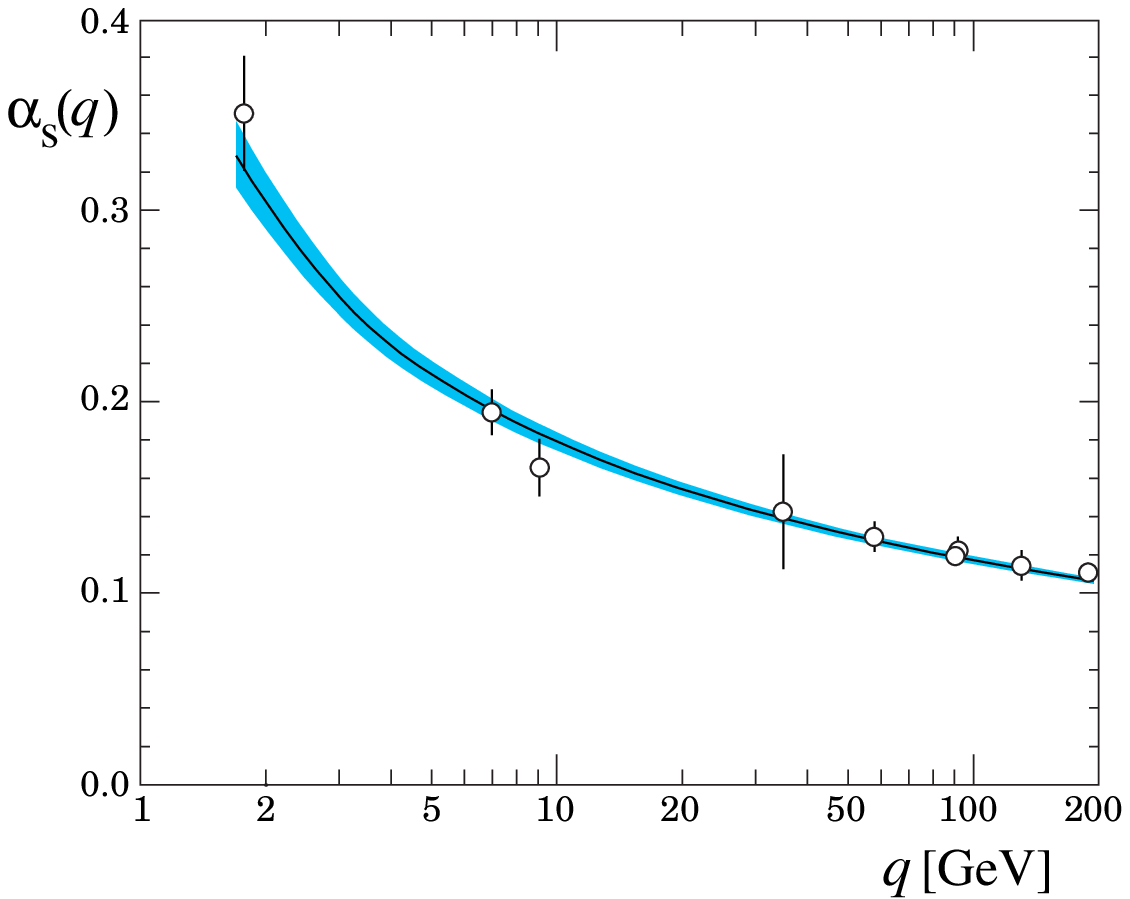}
\vskip0.4cm
\figurecaption{%
The experimentally measured values 
of the effective gauge coupling $\alpha_{\rm s}(q)$
confirm the theoretically expected behaviour [eq.~(1.3)]
at high energies
(compilation of the Particle Data Group [\ref{PDG}]).
}
\vskip0.0cm
}
\endinsert

The fact that the gauge coupling is weak at high energies
implies that perturbation theory will, in general, 
be reliable in this regime.
On the other hand, as we move towards the other end of the 
energy scale, the expansion breaks down at some point
and the physical picture associated with the Feynman diagrams
consequently becomes invalid. An immediate conflict between 
theory and the non-observation of isolated quarks is thus avoided.

\subsection 1.2 Lattice formulation

It remains to be shown, however, that 
quark confinement is indeed a property of QCD.
Moreover, once this is achieved, computational tools need to be 
developed to determine the basic properties of the 
quark bound states such as the pions, the kaons and the nucleons.

The lattice formulation of QCD was introduced many years ago 
as a framework in which these issues can be addressed
[\ref{Wilson}].
Introductory texts on the subject are
refs.~[\ref{BookI}--\ref{Handbook}], for example, 
and to find out about the most recent developments in the field,
the proceedings of the yearly lattice conferences 
usually provide a good starting point [\ref{LatticeConference}].

Very briefly lattice QCD is obtained by replacing the four-dimensional
space-time continuum through a hypercubic lattice and by
restricting the quark and the gauge fields to the lattice points.
The expression for the lagrange density (1.1) then needs
to be discretized in a sensible way, and  
the precise relation between 
the correlation functions calculated on the lattice
and the physical quantities of interest must be understood.
A few key elements of this construction are

\vskip1.2ex

\item{--}{%
The gauge symmetry can be fully preserved, and
no additional unphysical degrees of freedom are thus introduced.
}

\vskip1.2ex

\item{--}{%
In general the details of the discretization become irrelevant
in the continuum limit, i.e.~any reasonable lattice
formulation will give the same continuum theory up to finite
renormalizations of the gauge coupling and the quark masses.
}

\vskip1.2ex

\item{--}{%
Lattice QCD admits an expansion in Feynman diagrams that
coincides with the usual expansion up to terms proportional to 
the lattice spacing. The consistency of 
the lattice theory with the standard perturbative approach to QCD is thus 
guaranteed.
}

\vskip1.2ex
\noindent
A last point to be mentioned here 
is that the lattice provides a regularization
of the ultra-violet divergences that are 
usually encountered in quantum field theory.
The lattice theory is, therefore, mathematically well defined from the 
beginning.


\subsection 1.3 Numerical simulations

The application of numerical
simulation methods to solve the theory has been an interesting
perspective since the early days of lattice QCD. Today quantitative results
are practically all based on such numerical studies. 

In the course of these calculations the fields have to be stored 
in the memory of the computer, and only lattices
of limited size can thus be simulated. 
The progress in computer
technology allows the lattice extents to be doubled in all directions
roughly every 8 years. 
At present lattices as large as
$128\times64^3$ can be accommodated on 
(say) a commodity PC cluster with 128 nodes.
The simulation then proceeds by generating a representative ensemble
of fields through a stochastic process. Eventually
the physical quantities are extracted from ensemble 
averages of products of gauge-invariant local fields.

In general numerical simulations have the reputation of being
an approximate method that mainly serves to obtain qualitative
information on the behaviour of complex systems.
This is, however, not so in lattice QCD, 
where the simulations produce results that
are exact (on the given lattice) up to statistical errors. 
The systematic uncertainties related to the non-zero lattice spacing and
the finite lattice volume then still need to be investigated, but 
these effects are theoretically well understood and
can usually be brought under control.
 
\topinsert
\vbox{
\vskip0.0cm
\epsfxsize=6.6cm\hskip2.7cm\epsfbox{masses.eps}
\vskip0.3cm
\figurecaption{%
Recent calculations of the hadron masses $m_h$ (data points) in lattice
QCD agree quite well with the
observed spectrum (horizontal lines) [\ref{CPPACS}].
}
\vskip-0.0cm
}
\endinsert

Lattice QCD is being used to compute a wide range
of physical quantities, including the hadron mass spectrum (fig.~3),
decay constants, form factors and weak transition matrix elements.
The technique is not universally applicable, however,
and quantities like the inelastic proton-proton scattering cross-section
or the nucleon structure functions at small angles remain inaccessible.

\section 2. Quark confinement

As can be seen from fig.~3, there is currently no perfect match between 
experiment and 
the lattice calculations. The experts are not
alarmed by this, since the computations are still incomplete in certain
respects. However, rather than going into any details here, we now 
turn to the more fundamental question of quark confinement.

\subsection 2.1 Static quark potential

Quarks carry SU(3) quantum numbers and are thus
sources of the gauge field.
As a con\-sequence the latter exerts 
a force on the quarks, since any change in their positions
implies a change in field energy. According to the current
understanding, the confinement of 
quarks is associated with an unbounded increase of the field
energy at large quark separations.

The ground state energy of the gauge field hamiltonian
in the presence of a static 
quark-antiquark pair separated by a distance $r$
is referred to as the static quark potential $V(r)$.
There are various ways to compute $V(r)$ on the lattice.
Perhaps the most elegant approach is to consider 
Wilson lines that wind around 
a cylindrical space-time manifold (see fig.~4).
The correlation function of any two such loops
can be shown to satisfy
\equation{
  \langle P(r)P(0)^{\ast}\rangle
  \mathrel{\mathop=_{T\to\infty}}
  \rme^{-TV(r)}\left\{1+\rmO\bigl(\rme^{-T\epsilon}\bigr)\right\},
  \qquad\epsilon>0,
  \enum
}
where $T$ denotes the circumference of the cylinder, and
the potential can thus be determined
by calculating the correlation function.

\topinsert
\vbox{
\vskip0.0cm
\epsfxsize=5.0cm\hskip4.0cm\epsfbox{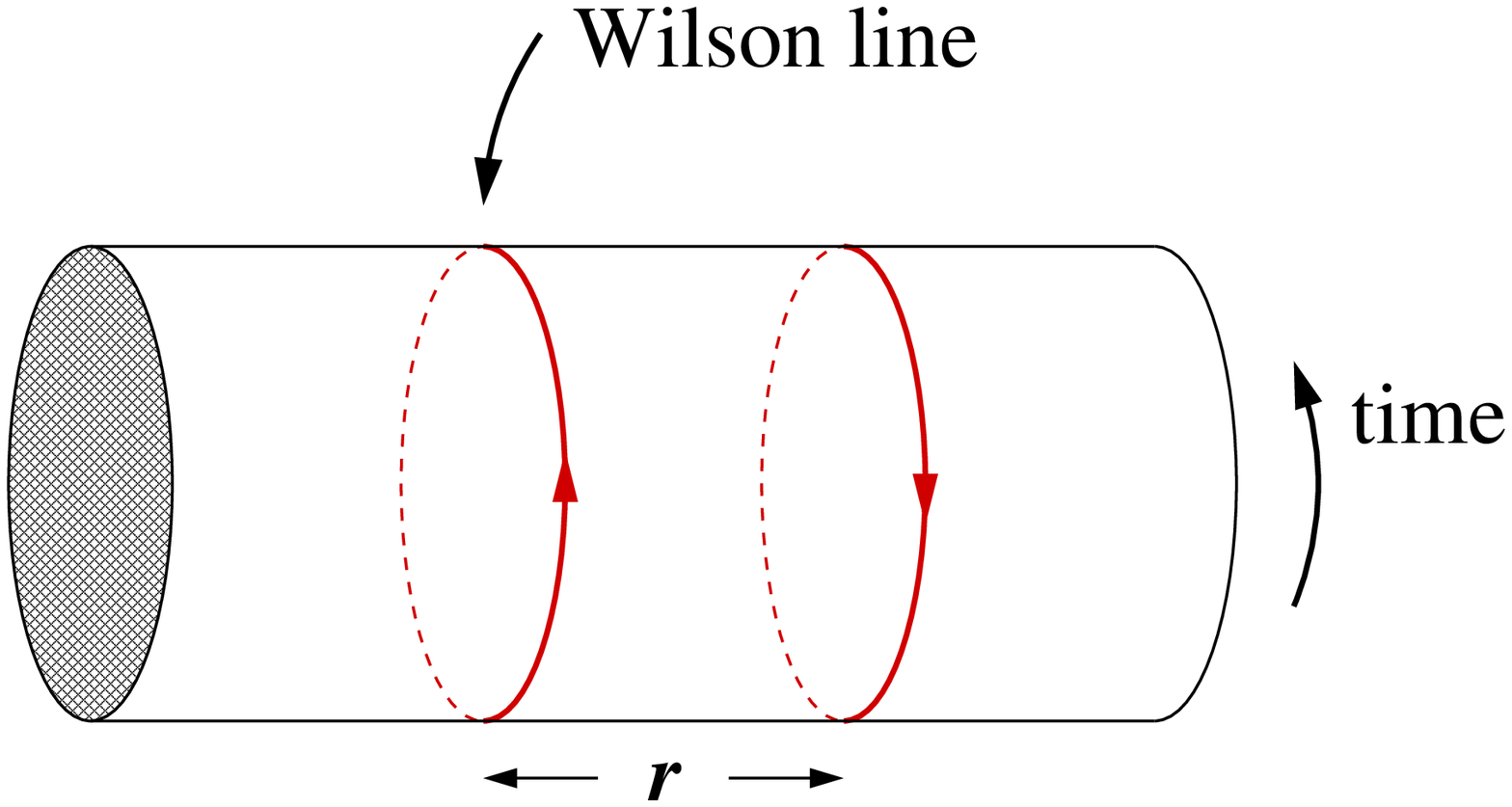}
\vskip0.4cm
\figurecaption{%
Polyakov loops are Wilson lines that wind around the
space-time manifold in the time direction. In eq.~(2.1)
we consider the correlation function of 
two Polyakov loops with opposite orientation that are
separated by a distance $r$ in space.
}
\vskip0.0cm
}
\endinsert

Some recent results for the potential $V(r)$ and the force $V'(r)$ are
plotted in fig.~5. They show that $V(r)$ is monotonically rising and
eventually grows linearly with a small correction proportional to 
$1/r$. The force is fairly strong, at least $1\,\GeV/\fm$ in the 
whole range of distances, and 
if this continues to be so at larger values of $r$
it will evidently not be possible to separate the quark-antiquark pair. 
At short distances, on the other hand,
the data points rapidly approach the curves that are obtained 
in perturbation theory (as it has to be
since the effective gauge coupling is small in this regime). 

\topinsert
\vbox{
\vskip0.0cm
\epsfxsize=6.0cm\hskip-0.3cm\epsfbox{pot.eps}%
\epsfxsize=5.7cm\hskip0.9cm\epsfbox{fr.eps}
\vskip0.4cm
\figurecaption{%
Simulation results for the static potential $V(r)$ [\ref{SilviaRainer}]
and the force $V'(r)$ [\ref{MultiLevel}]. 
The statistical and systematic
errors in these calculations are smaller than or at most equal
to the size of the data symbols. 
}
\vskip0.2cm
}
\endinsert

\subsection 2.2 String model

The data shown in the second plot in fig.~5
suggest that the force $V'(r)$ 
approaches a constant $\sigma\simeq1.06\,\GeV/\fm$ 
at large quark-antiquark separations.
$\sigma$ is referred to as the string tension, 
because a constant force is 
what would be obtained if the quarks were held together 
by an elastic string.

The idea that quark confinement is linked to the formation of 
string-like flux tubes has in fact been around for very many years.
We should then not only see the linear rise in the static potential
but also a characteristic $1/r$ correction that derives from the 
zero-point energy of the transversal
string vibrations. Explicitly the prediction is that
[\ref{WKB},\ref{UniversalTerm}]
\equation{
  V(r)\mathrel{\mathop=_{r\to\infty}}
  \sigma r+\mu-{\pi\over12r}+\rmO(r^{-2}),
  \enum
}
where (from the string theory point of view)
$\sigma$ and the mass $\mu$ are free parameters.

The string picture may appear to be somewhat naive, but 
the simulation data for the 
second derivative of the static potential
shown in fig.~6 agree very well with eq.~(2.2).
What is plotted there is the dimensionless combination
$-\frac{1}{2}r^3V''(r)$, which, according 
to string theory, should converge to $\pi\over12$ 
at large $r$.
If we allow for a small higher-order 
correction proportional to $1/r$, 
the data are perfectly compatible with this.

\topinsert
\vbox{
\vskip0.0cm
\epsfxsize=9.2cm\hskip2.0cm\epsfbox{cr3d.eps}
\vskip0.4cm
\figurecaption{%
Simulation results for the second derivative of the static potential
in $d=4$ space-time dimensions (upper curves and data points) 
and in three dimensions [\ref{MultiLevel}].
}
\vskip0.0cm
}
\endinsert

An even better matching between string theory and the gauge theory
is observed in three space-time dimensions. In this case a string 
with fixed ends can only
vibrate in one transversal direction and the associated
zero-point energy is consequently reduced by a factor $2$.
This is exactly what one finds on the lattice when the three-dimensional
gauge theory is simulated (points and curves in the lower half
of fig.~6).

\subsection 2.3 Summary

The lattice studies conducted so far leave little doubt that the 
quarks are confined because the field energy of the surrounding SU(3) 
gauge field rises linearly at large distances.
As far as the static potential is concerned,
we have also seen that
the behaviour of the gauge theory in this regime 
is accurately described by an effective string model.
This correspondence is, incidentally, expected to extend 
to the confinement phase of other gauge field theories
[\ref{CaselleEtAl}--\ref{CaselleEtAlII}]
and to a range of other observables such as 
the spectrum of excited states in the presence
of static quarks 
[\ref{MichaelPerantonis},\ref{KutiEtAl}].

\vfill\eject

\section 3. QCD from low to high energies

As explained above, lattice QCD allows us to study 
the phenomenon of quark confinement and to compute 
the basic properties of the light mesons and of the nucleons.
Among the most obvious quantities to consider in this low-energy
regime are the
mass $\mpi$ of the charged pions and the pion decay constant $\fpi$.
Experimentally the latter is determined by measuring the decay rate
$\Gamma_{\pi^{+}\to\mu^{+}\nu_{\mu}}$, while 
in QCD it can be expressed as a matrix element
\equation{
  \langle0|A_{\mu}|\pi^{+}\,p\rangle=ip_{\mu}f_{\pi}
  \enum
}
of the appropriate axial quark current $A_{\mu}$
between the vacuum state and the one-pion state with four-momentum $p$.

The experiments at the big particle colliders, on the other hand, 
probe the interactions of the quarks and gluons at high energies
(from 10 to 100 and more GeV),
where asymptotic freedom has set in and perturbation theory 
may be applied to calculate the reaction rates.
The hadronic decay width of the $Z$--boson, for example, is
given by
\equation{
  \Gamma_{Z\to q\bar{q}}=\hbox{constant}\times
  \left\{1+{\alpha_{\rm s}\over\pi}+\rmO(\alpha_{\rm s}^2)\right\},
  \enum
}
where $\alpha_{\rm s}$ denotes the gauge coupling (1.3) at momentum 
transfer $q$ equal to the mass of the $Z$--boson
and the constant includes the (calculable) contributions of 
the weak and electromagnetic interactions.

\subsection 3.1 Connecting different energy regimes

From a purely phenomenological point of view, 
it seems unlikely that there is any connection 
between (say) the pion mass and the $Z$--boson 
decay rate. However, since all strong interaction physics
is described by the same underlying field theory,
there have to be at least some such relations.

To make this a bit more explicit, first note that 
in QCD any physical quantity is a function of the 
parameters that appear in the lagrangian (1.1).
The gauge coupling $g$ is one of them, 
but for the following discussion it is actually more natural
to consider the scale $\Lambda$ in eq.~(1.3) 
to be a basic parameter of the theory.
We then infer that there are functions $G$ and $F$ such that
\equation{
  \mpi^2/f_{\pi}^2=G(m_u/\Lambda,m_d/\Lambda,\ldots),
  \enum
  \next{2ex}
  \fpi/\Lambda=F(m_u/\Lambda,m_d/\Lambda,\ldots),
  \enum
}
and for the masses of the kaons and the heavier pseudo-scalar mesons
similar equations can be written down.

Now if we take the experimental values of
the meson masses in units of the pion decay constant as input,
the equations for the meson masses can be solved for the quark masses 
(in units of $\Lambda$)
and the ratio $\fpi/\Lambda$ then becomes a calculable quantity.
Note that this ratio links the physical
properties of the pions to the $\Lambda$ parameter,
which is a characteristic scale in the high-energy regime of QCD.
Since $\fpi/\Lambda$ is also known experimentally,
an interesting test is thus obtained that 
will only be passed (barring miracles) if QCD is the correct theory at 
all energies.

\subsection 3.2 The scale problem

In lattice QCD the calculation of $\fpi/\Lambda$ 
appears to be a difficult task,
because $\fpi$ and $\Lambda$ are physical quantities
that refer to the properties of the theory 
at energies orders of magnitude apart.
A straightforward approach then requires the simulation of 
lattices with a very large number of points.

It is not difficult to obtain an estimate of what would be needed.
For the computation of $\fpi$ and the meson masses, 
the spatial lattice size $L$
should be at least $2\,\fm$ as otherwise there
will be sizeable finite-volume effects.
The effective gauge coupling $\alpha_{\rm s}(q)$, on the other hand, 
can only be reliably determined at momenta $q$ up to $1/a$ 
or so (where $a$ denotes the lattice spacing). 
Moreover, to able to extract $\Lambda$ from the 
asymptotic behaviour (1.3) of the coupling,
$q$ must be taken to values deep in the high-energy regime of QCD.
The combination of all these requirements then
implies that the number $L/a$ of lattice
sites in each direction has to be on the order of 100 or 
maybe even larger than this.

\subsection 3.3 Finite-size scaling

Such lattices may become accessible
at some point, but it is much more efficient to adopt a recursive scheme,
where the large scale difference is bridged by a sequence of 
matching lattices [\ref{Scaling}] 
(see ref.~[\ref{LesHouches}] for an introduction to the subject). 
The key idea is to introduce an effective gauge coupling
$\alpha(q)$ that measures the interaction strength
at a momentum $q$ proportional to $1/L$. In this way
the finite lattice size becomes a device to probe the interactions
rather than being a source of systematic errors.

There are many ways to define an effective coupling of 
this kind.
We may choose some particular boundary conditions, for example, 
and take the response of the system
to a change in the boundary values of the fields 
as a measure for the interaction 
strength [\ref{ALPHAPureGaugeI}]. The important point to note is that
the final results (such as $\fpi/\Lambda$) do not depend
on any of these details.

\topinsert
\vbox{
\vskip0.0cm
\epsfxsize=10.0cm\hskip1.0cm\epsfbox{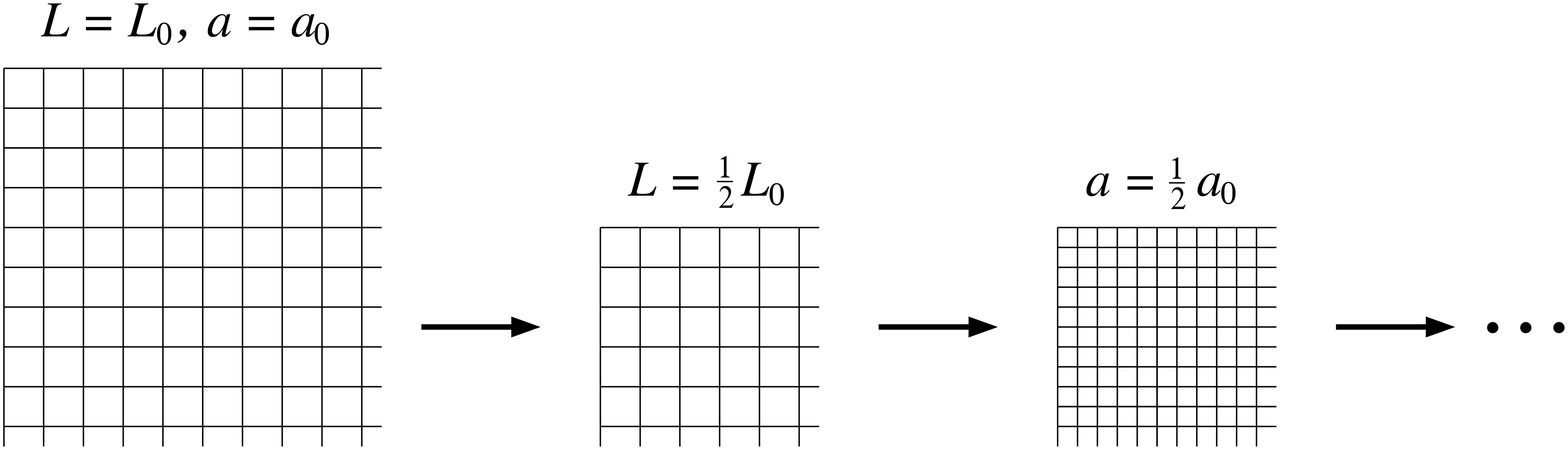}
\vskip0.4cm
\figurecaption{%
Finite-size scaling amounts to the construction of a sequence of 
matching lattices with lattice spacings $a=2^{-k}a_0$
and decreasing physical sizes $L$.
}
\vskip0.0cm
}
\endinsert

The next step is to construct a sequence of lattices
with lattice spacings $a$ and 
sizes $L$ that decrease by factors of $2$ as shown in fig.~7. 
Starting with some physically large lattice,
it is possible to scale the lattices
to very small sizes in this way without running into
technical difficulties.  
The coupling $\alpha(q)$ (the one discussed above 
that is defined at some $q$ proportional to $1/L$)
can be calculated on each of these lattices
and its evolution can thus be followed over a 
wide range of momenta.

\subsection 3.4 Simulation results

So far such computations have been performed
in the pure gauge theory 
and in QCD with two flavours of quarks.
As can be seen from fig.~8,
the momentum dependence of
the coupling $\alpha(q)$ that has been studied
is accurately matched by perturbation theory already at
fairly low momenta.
Contact with the asymptotic behaviour (1.3) of the coupling
can thus easily be made, and since also the low-energy regime
is safely reached, these calculations provide the desired link
between that regime and the $\Lambda$ parameter.

\topinsert
\vbox{
\vskip0.0cm
\epsfxsize=6.6cm\hskip-0.2cm\epsfbox{alphaSF_Nf0.eps}%
\hskip0.1cm\epsfxsize=5.83cm\epsfbox{alphaSF_Nf2.eps}
\vskip0.4cm
\figurecaption{%
Momentum-dependence of the effective gauge coupling 
(in a particular finite-volume scheme) in the pure SU(3) gauge theory
[\ref{ALPHAPureGaugeI}--\ref{ALPHAPureGaugeIII}]
and in QCD with two flavours of massless quarks
[\ref{ALPHALight}].
}
\vskip0.0cm
}
\endinsert

For the full theory, with all flavours of quarks properly
included, a similar study has not yet been made.
The reason partly is that these calculations are 
exceedingly expensive in terms of computer time,
because the available simulation techniques 
become very inefficient once the quark polarization effects
are taken into account (fig.~9).
We shall return to this issue in a moment and only note
at this point that 
practically all applications of lattice QCD
meet the same difficulty.

Since they are often not very large, it is,
however, a sensible and still widely used approximation to neglect the
quark polarization effects.
The calculations of the pion decay constant
and of the pseudo-scalar meson masses that are required to 
obtain $\fpi/\Lambda$ can then be carried out with the 
presently available computer resources
(see refs.~[\ref{CPPACS},\ref{ALPHAUKQCD}] for example).
As already mentioned,
the lattices should be at least $2$ fm wide in these computations, but
this is not a problem here because
the lattice spacing does not need to be extremely small 
at the same time.
Values from $0.05$ to $0.1$ fm have actually been found 
to be adequate in this context.

Once all this is done,
and a careful analysis of the systematic errors is made,
the combination of the results yields the value
[\ref{ALPHAPureGaugeII},\ref{ALPHAUKQCD}]
\equation{
  \fpi/\Lambda=0.56\pm0.05.
  \enum
}
This figure agrees with the experimental number $\fpi/\Lambda=0.62\pm0.10$
within the quoted errors, which is somewhat unexpected since 
the quark polarization effects
have been neglected. Maybe the ratio is not strongly 
affected by them,
but there is currently no very good theoretical argument for this and
the coincidence can, therefore, not be taken as a solid confirmation of QCD
at this point.
The computation nevertheless provides important insights into how
precisely the low- and the high-energy regimes of the theory are
connected to each other,
and it also shows the potential of the lattice approach 
to lead, in due time, to some very stringent tests of QCD.

\topinsert
\vbox{
\vskip0.0cm
\epsfxsize=2.5cm\hskip4.5cm\epsfbox{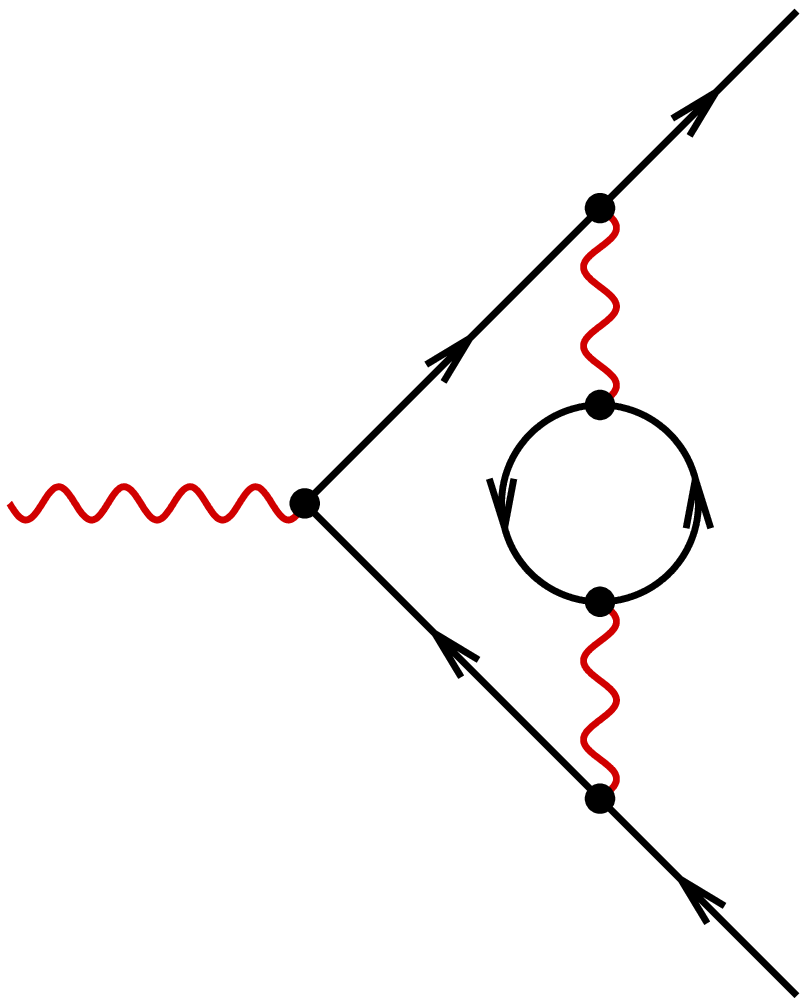}
\vskip0.2cm
\figurecaption{%
In the language of Feynman diagrams, quark polarization 
effects are represented by subdiagrams with closed quark lines. 
For technical reasons,
many results in lattice QCD still do not include these contributions.
}
\vskip0.2cm
}
\endinsert

\subsection 3.5 Real-world lattice QCD simulations

At present the great challenge in lattice QCD is
to devise efficient simulation methods for the full theory
(including quark polarization effects)
that will work well on large lattices and at small quark masses.
The fact that the currently available techniques 
will not lead very far
became particularly clear 
at the lattice conference last year in Berlin 
[\ref{LatticeConference}],
where an effort was made to assess the cost of such calculations.

The figure that is usually quoted in this context is 
the number of arithmetic operations that are required
to generate the next 
statistically independent field configuration.
An approximate and mostly empirical formula that was presented
at the conference is [\ref{Ukawa}]
\equation{
 {\hbox{\ninepoint \# operations}\over\hbox{\ninepoint field configuration}}
 \simeq 3.3\left[{140\,\MeV\over\mpi}\right]^6
 \left[{L\over 3\,\fm}\right]^5\left[{0.1\,\fm\over a}\right]^7
 \hbox{Tflops year},
 \enum
}
which shows the dependence on the lattice parameters
and on the calculated value $\mpi$ of the pion mass (which depends on
the specified values of the quark masses).
The result is given in Tflops years, 
the number of operations that a computer
with a sustained speed of $1$ Tflops 
($10^{12}$ floating-point operations per second) performs in
$1$ year of running time. 

While such machines will become available to the lattice 
community in the near future, the poor scaling behaviour of the 
algorithms (the high powers in the formula) tells us
that it will not be possible to vary the lattice parameters 
over a wide range. If the lattice spacing is decreased by a 
factor of $2$, for 
example, the simulation time goes up 
by a factor of $128$ or so.
Evidently this unfavourable situation calls for new algorithmic ideas,
and a significant investment in purely technical R\&D
work will be required in the coming years to solve the problem.

\section 4. Conclusions

Lattice QCD was introduced nearly 30 years ago and it has since then
turned into a powerful quantitative approach to the physics
of the strongly interacting particles. Only 
a few topics have been touched in this talk, 
but this should not hide the fact
that the technique is being used to calculate many quantities
of phenomenological interest (hadron masses, decay constants, 
transition matrix elements, and so on)
and also to study some of the  more fundamental issues 
such as the breaking of chiral symmetry and
the phase diagram at non-zero baryon density.

All this work has had its share in making QCD the unique candidate
of the theory of the strong interactions. 
In particular, little doubt is left that the non-linearities 
of the dynamics of the SU(3) gauge field are responsible for 
the confinement of the quarks.
Moreover we have seen that 
quark confinement and 
asymptotic freedom are just two complementary
aspects of the theory.
The fact that the relation between the parameters in the lagrangian
and the basic properties of the mesons and nucleons
can be worked out on the lattice is the key to showing this,
along with the ability to compute the (non-perturbative) evolution of the 
effective gauge coupling from very low to high energies.

Currently the quark polarization effects are 
often neglected in the numerical simulations because 
their inclusion slows down the computations by a large factor.
It is mainly for this reason that comparisons of the lattice results
with experimental numbers cannot at present be regarded as hard tests of QCD.
Once this technical limitation is 
overcome, it is clear, however, that precision tests will become
a reality, and the hope of the experts is that this will happen
before the next 30 years have elapsed!

\vskip1ex

I am indebted to Rainer Sommer for sending a set of data tables 
and to Peter Weisz and Hartmut Wittig for helpful discussions and
correspondence.
Many thanks also go to
Daniel Iagolnitzer and Jean Zinn-Justin for having
organized this unique conference.

\beginbibliography


\bibitem{GrossWilczek}
D. J. Gross, F. Wilczek,
Phys. Rev. Lett. 30 (1973) 1343

\bibitem{Politzer}
H. D. Politzer,
Phys. Rev. Lett. 30 (1973) 1346


\bibitem{PDG}
C. Caso et al. (Particle Data Group), Eur. Phys. J. C3 (1998) 1

\bibitem{Bethke}
S. Bethke,
J. Phys. G26 (2000) R27


\bibitem{Wilson}
K. G. Wilson, Phys. Rev. D10 (1974) 2445


\bibitem{BookI}
I. Montvay, G. M\"unster,
Quantum fields on a lattice
(Cambridge University Press, Cambridge, 1994)

\bibitem{BookII}
H. J. Rothe,
Lattice gauge theories: an introduction, 2nd ed.
(World Scientific, Singapore, 1997)

\bibitem{Rajan}
R. Gupta,
Introduction to lattice QCD,
in: Probing the
standard model of particle interactions (Les Houches 1997),
eds. R. Gupta et al. 
(Elsevier, Amsterdam, 1999)

\bibitem{Handbook}
A. Kronfeld,
Uses of effective field theory in lattice QCD,
in: Handbook of QCD, Vol. 4, ed. M. Shifman (to appear),
hep-lat/0205021


\bibitem{LatticeConference}
M. M\"uller-Preussker et al. (eds.),
Proceedings of the
19th International Symposium on Lattice Field Theory,
Nucl. Phys. B (Proc. Suppl.) 106 \& 107 (2002)


\bibitem{CPPACS}
S. Aoki et al. (CP-PACS Collab.),
Light hadron spectrum and quark masses from quenched lattice QCD,
hep-lat/0206009


\bibitem{SilviaRainer}
S. Necco, R. Sommer,
Nucl. Phys. B622 (2002) 328

\bibitem{MultiLevel}
M. L\"uscher, P. Weisz,
J. High Energy Phys. 07 (2002) 049


\bibitem{WKB}
M. L\"uscher, K. Symanzik, P. Weisz,
Nucl. Phys. B173 (1980) 365

\bibitem{UniversalTerm}
M. L\"uscher,
Nucl. Phys. B180 (1981) 317

\bibitem{CaselleEtAl}
M. Caselle, R. Fiore, F. Gliozzi, M. Hasenbusch, P. Provero,
Nucl. Phys. B486 (1997) 245

\bibitem{Majumdar}
P. Majumdar,
Experiences with the multilevel algorithm,
hep-lat/0208068 

\bibitem{CaselleEtAlII}
M. Caselle, M. Hasenbusch, M. Panero,
String effects in the 3d gauge Ising model,
hep-lat/0211012

\bibitem{MichaelPerantonis}
S. Perantonis, C. Michael,
Nucl. Phys. B347 (1990) 854

\bibitem{KutiEtAl}
K. J. Juge, J. Kuti, C. J. Morningstar,
Fine structure of the QCD string spectrum,
hep-lat/0207004


\bibitem{Scaling}
M. L\"uscher, P. Weisz, U. Wolff,
Nucl. Phys. B359 (1991) 221

\bibitem{LesHouches}
M. L\"uscher,
Advanced lattice QCD,
in: Probing the
standard model of particle interactions (Les Houches 1997),
eds. R. Gupta et al.
(Elsevier, Amsterdam, 1999)


\bibitem{ALPHAPureGaugeI}
M. L\"uscher, R. Sommer, P. Weisz, U. Wolff,
Nucl. Phys. B413 (1994) 481

\bibitem{ALPHAPureGaugeII}
S. Capitani et al. (ALPHA collab.),
Nucl. Phys. B544 (1999) 669

\bibitem{ALPHAPureGaugeIII}
Jochen Heitger et al. (ALPHA collab.),
Nucl. Phys. B (Proc. Suppl.) 106 (2002) 859

\bibitem{ALPHALight}
Achim Bode et al. (ALPHA collab.),
Phys. Lett. B515 (2001) 49


\bibitem{ALPHAUKQCD}
J. Garden et al. (ALPHA \& UKQCD collab.),
Nucl. Phys. B571 (2000) 237




\bibitem{Ukawa}
A. Ukawa (CP-PACS \& JLQCD collab.),
Nucl. Phys. B (Proc. Suppl.) 106 (2002) 195

\endbibliography

\bye